\documentclass{mem}
\usepackage{natbib}\usepackage{txfonts}\usepackage{balance}
\usepackage{graphicx}
\usepackage{hyperref}
\usepackage{float}
\idline{75}{282}

\newcommand{\msun}{M$_\odot$~}
\newcommand{\feh}{[Fe/H]}
\newcommand{\parsec}{\texttt{PARSEC}~}

\begin{document}
\def\teff{$T\rm_{eff }$}
\def\kms{$\mathrm {km s}^{-1}$}

\title{
From the cosmological Li problem to the Galactic Li evolution
}

   \subtitle{}

\author{
Xiaoting Fu\inst{1} 
          }

\institute{
KIAA-The Kavli Institute for Astronomy and Astrophysics at Peking University, Beijing 100871, China.
\email{xiaoting.fu@pku.edu.cn}
}

\authorrunning{Fu}

\titlerunning{Li problem and the Galactic evolution}

\abstract{
Li is the very element that presents deep insights yet many problems to astrophysics. 
In this proceeding manuscript I first introduce a stellar solution to the cosmological Li problem,
which reveals that Li was first destroyed and re-accumulated by these stars shortly after they were born,
then discuss Li in the interstellar medium in the form of the molecule LiH,
and the different Li enrichment histories seen in the Galactic thick and thin discs with data from the $Gaia$-ESO survey.  
}
\maketitle{}

\section{Introduction}

\begin{figure*}[h]
\resizebox{\hsize}{!}{\includegraphics[clip=true]{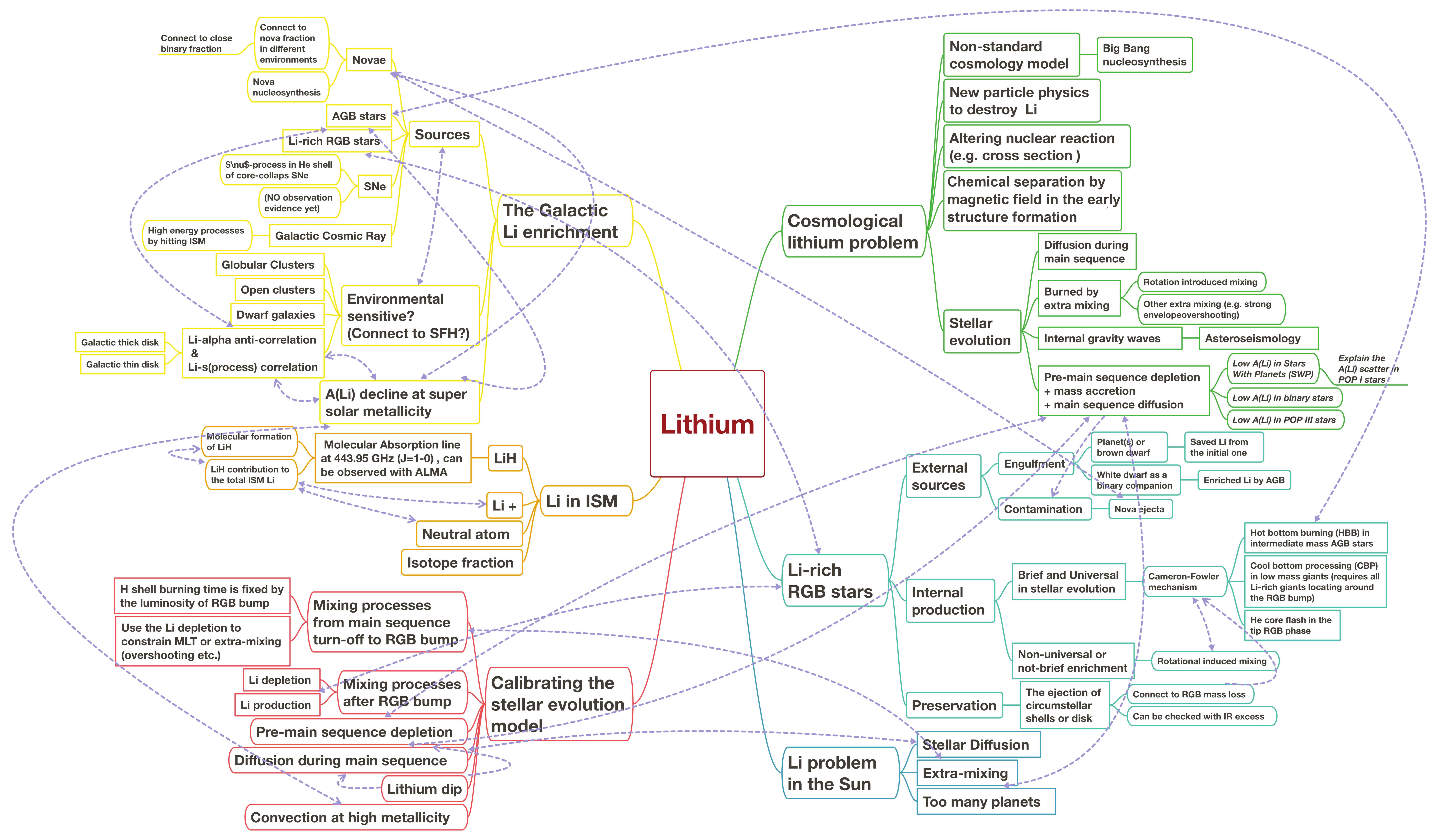}}
\caption{\footnotesize
Mind map on Li studies in astrophysics.}
\label{limap}
\end{figure*}

Lithium ($^7$Li) is the very element that presents deep insights yet many problems to astrophysics.
It is a primordial element from the Big Bang nucleosynthesis (BBN) and is considered a tester to the standard BBN theory.
Because it is easily burned via the reaction $^7Li(p,\gamma)^4He+^4He$~ at temperature as low as several $10^6$ K,
its stellar abundance is also a probe to investigate the stellar structure, 
especially to study the stellar interior convection which greatly affect stellar nucleosynthesis and thereafter chemical evolution in the galaxy.
However, many problems on Li have puzzled (also inspired) the community for decades.
Here I present a mind map in Fig. \ref{limap}, illustrating the Li studies in astrophysics and 
how they link different fields of researches
(i.e. cosmology, nuclear physics, stellar physics, molecular chemistry, high energy physics, planetary Science, Galaxy evolution).
The Li-raised problems have plagued our understanding of these fields.

In this proceeding manuscript, 
I will focus on the cosmological Li problem (Section \ref{sec:pms}),
the molecular Li in the interstellar medium (Section \ref{sec:ism}),
and the Galactic Li evolution seen in the thick and thin disc stars (Section \ref{sec:ges}).

\section{A stellar solution to the Cosmological Li problem}
\label{sec:pms}

The known \textit{``cosmological lithium problem''} refers to 
the discrepancy that the BBN theory predicts a primordial Li abundance three times higher
than observed in old metal-poor stars.
As illustrated in Fig. \ref{limap}, different fields of study have been pursued for possible solutions to the problem,
many of them are being presented in this conference.
To seek for the missing Li that makes the gap between the BBN prediction and the observed Li in the old metal-poor stars,
in \citet{fu2015} we propounded a stellar model to solve the problem.

\begin{figure}[h!]
\resizebox{\hsize}{!}{\includegraphics[clip=true]{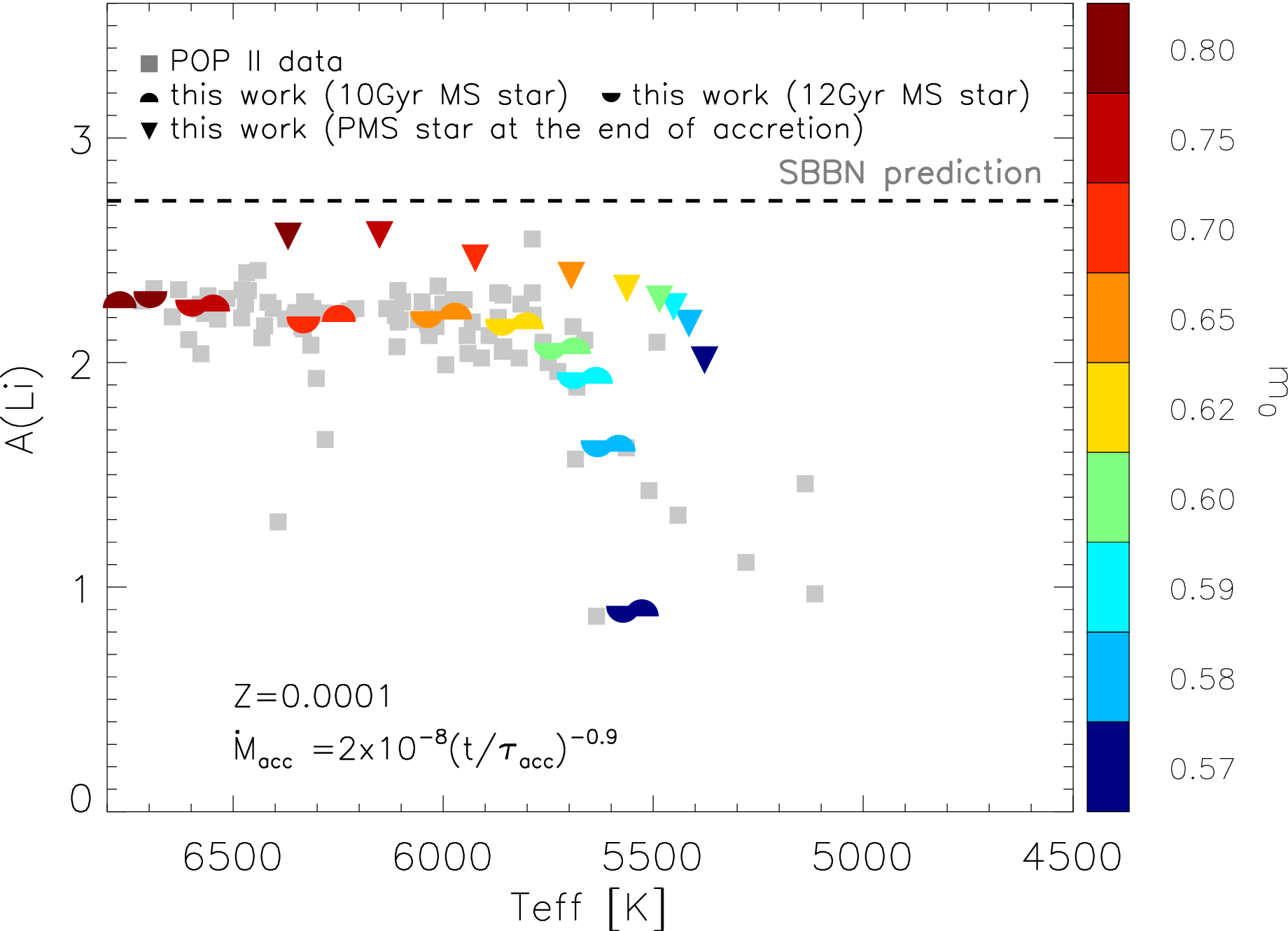}}
\caption{\footnotesize
        Model results in comparison with Li measurements collected in \citet[][ grey filled squares]{molaro2012}. 
        The filled triangles mark the model predictions at the end of the late accretion.
        The filled upper circles and lower circles are main sequence A(Li) at 10 Gyr and 12 Gyr.
        Symbols are colour-coded with the initial stellar mass.
        The black dashed line is the prediction by standard BBN.
        }
\label{fig:pms}
\end{figure}

We successfully reproduce the observed Li plateau and the declining branch at low temperatures,
with an initial Li abundance of the BBN prediction value.
Fig. \ref{fig:pms} shows the model results.
The age range, 10 Gyr (filled upper circles) to 12 Gyr (filled lower circles), is the typical age of metal-poor stars in the Galactic halo.
Our stellar model is based on the stellar evolution code \parsec \citep{bressan2012}.
In this set of model we take into account the  pre-main sequence (PMS) evolution including
effects of convective overshooting (OV) and residual mass accretion.  
During the  main sequence (MS) phase, the conventional nuclear burning and microscopic diffusion are considered as in the standard \parsec models.
We find that $^7$Li could be significantly depleted by convective OV in the PMS phase and then partially restored in the stellar atmosphere by a tail of matter accretion. 
The residual accretion  could be regulated by EUV photoevaporation from the star itself. 
The whole process is self-regulated by the stellar mass.

\begin{figure}
\resizebox{\hsize}{!}{\includegraphics[clip=true]{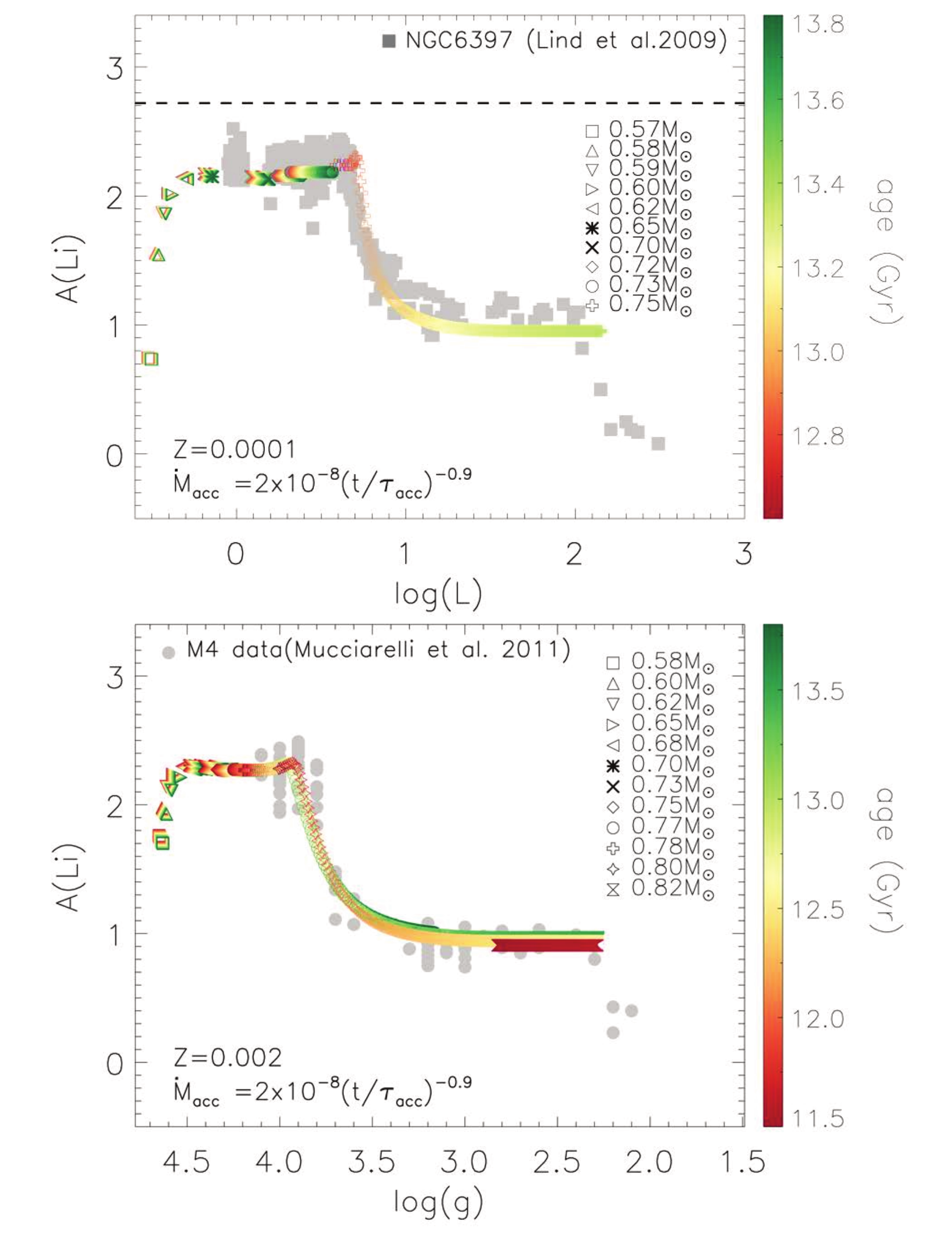}}
\caption{\footnotesize
        $Upper~panel$: A(Li) as a function of log(L) for stars with mass 0.57–0.75 \msun between 12.6 Gyr and 13.8 Gyr, 
        in comparison with  NGC\,6397 \citep{lind2009} data.
        $Lower~panel$:  A(Li) vs. log(g) for stars with mass 0.58–0.82 \msun between 11.5 Gyr and 13.8 Gyr.
         M4 measurements \citep{mucciarelli2011} are the grey filled dots.
        In both panel the stellar age is color-coded. }
\label{gc}
\end{figure}

With the same set of model, we not only close the gap between the observed Li plateau and the BBN prediction,
but also well reproduce Li abundance observed in the Galactic Globular Clusters.
Fig. \ref{gc} display our model predictions for NGC\,6397 and M4 in comparison with the observed values.

\section{LiH in the ISM}
\label{sec:ism}

To avoid the uncertainties from star formation process and stellar evolution, 
the most direct way to investigate the initial Li abundance is observing Li in ISM. 
Molecular studies suggest that the majority of the interstellar Li is in the form of lithium hydride (LiH), 
which is well mixed in the molecular gas phase \citep{cw1998}.

\begin{figure}
\resizebox{\hsize}{!}{\includegraphics[clip=true]{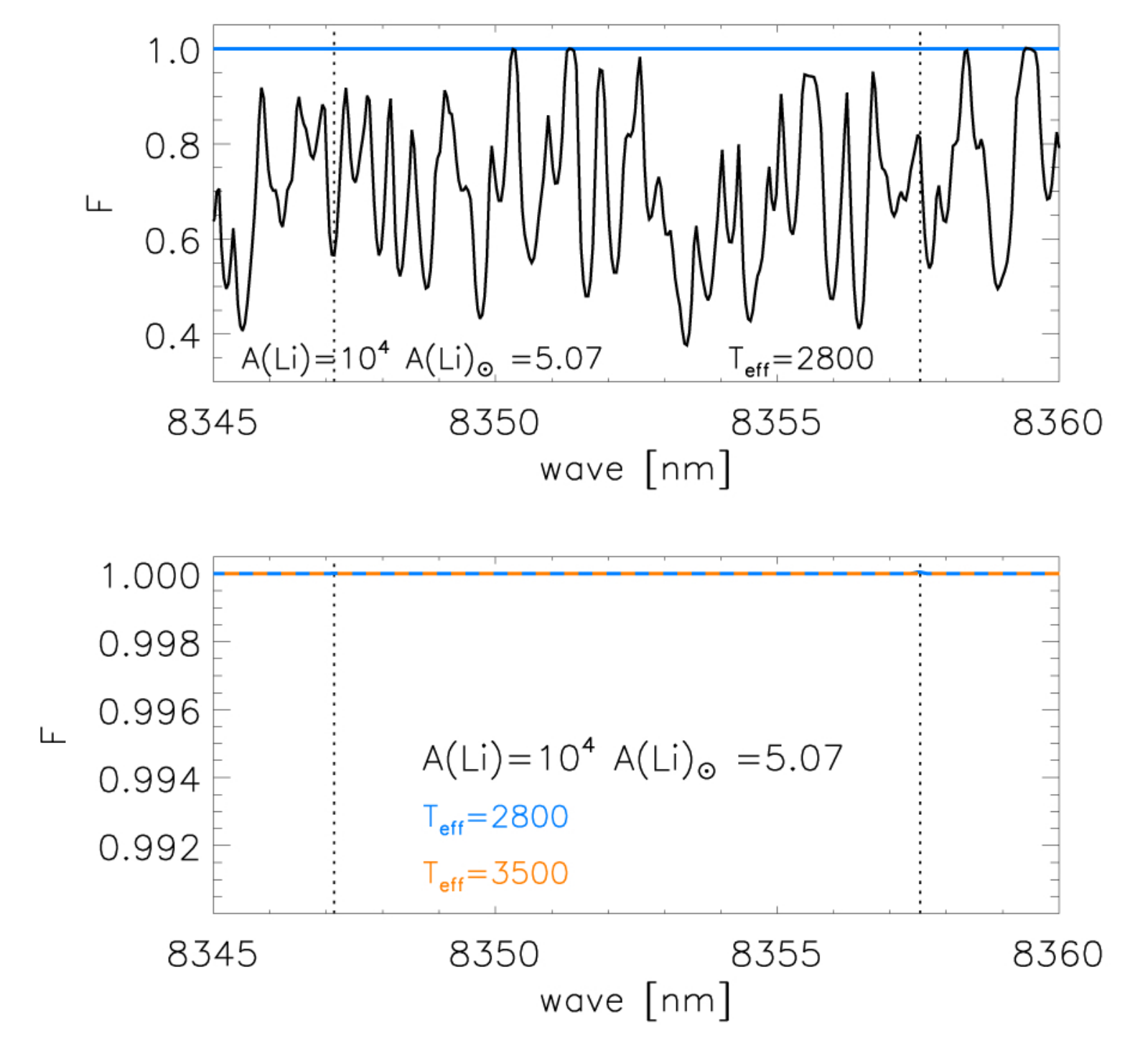}}
\caption{\footnotesize
        Synthetic stellar spectra around the 8347\,nm and 8357\,nm LiH lines' region with A(Li)=5.07.
        The two vertical dashed lines mark the LiH line wavelength. 
$Upper~panel:~$ With all species of molecular and atomic lines.
$Lower~panel:~$ With only LiH line.
The synthetic spectra are calculated with the COMA code \citep{Aringer2016} and provided by Bernhard Aringer.}
\label{lih_nir}
\end{figure}

Fig. \ref{lih_nir} shows two stellar synthetic near infrared (NIR) spectra with an extremely high Li abundance of A(Li)=5.07.
Though there are LiH lines around 8347\,nm and 8357\,nm in infrared,
they are ultra weak and are strongly blended with other molecular lines nearby.
For the ISM spectra in NIR, the lines will be even more difficult to measure.
Thus the only possible line to measure LiH is the J=1-0 line at 443.95 GHz (675 $\mu m$).

The transmission curve on the ground around this frequency is very low and makes it very difficult to observe.
There are measurements of LiH upper limit reported in the literature \citep{cw1998, friedel2011}.
Recently we have ALMA  LiH (J=1-0)  observations in the Milky Way toward two interstellar clouds.
Even with the best weather and best condition of ALMA, we could only get marginal LiH detections.
The result abundance is lower than the Galactic chemical evolution model predictions, 
which indicate that LiH might not be the main form of Li in the ISM.

Going to space is the only left way to seek for LiH.
The sub-THz instrument \citep{csst} on China Space Station Telescope (CSST) that is planned to be launched around 2025
may shed a new light on this problem.

\section{The Galactic Li evolution in the thick and thin discs}
\label{sec:ges}

Though interstellar Li abundance is very difficult to obtain,
stellar Li abundance could be an index of the Galactic Li evolution.

\begin{figure}
\resizebox{\hsize}{!}{\includegraphics[clip=true]{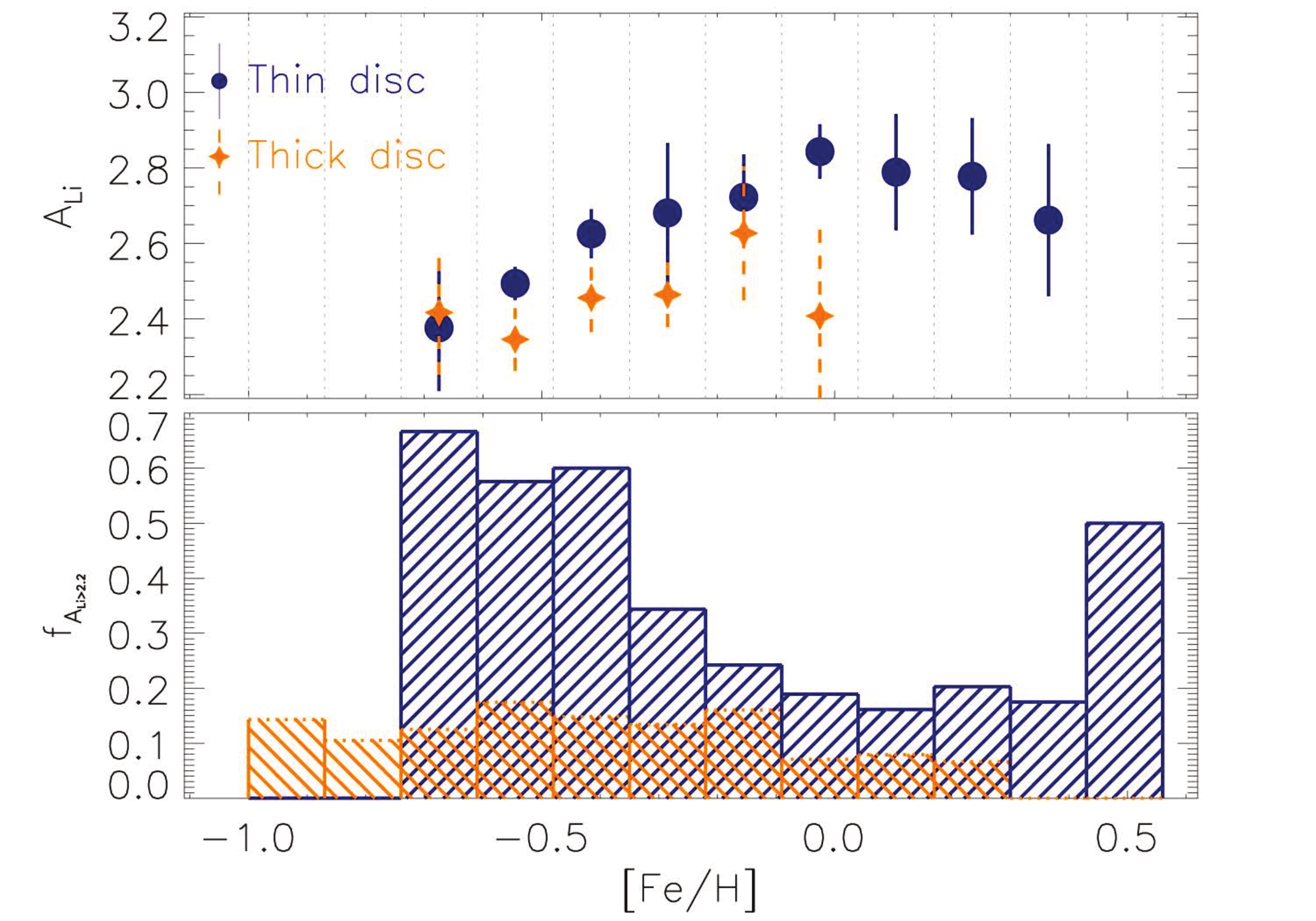}}
\caption{\footnotesize
        Stars with enriched Li (A(Li) > 2.2 dex). 
        $Upper~panel$:  mean A(Li) values (error-weighted) of the six stars with the highest Li abundance in each \feh~ bin.
        Blue filled dots represent the thin disc and the orange stars indicate the thick disc. 
        $Lower~panel$: fractions of Li-enriched stars ( $f_{A(Li)}>2.2$) in each \feh~ interval 
        for thin disc stars (blue-shades histogram) and thick disc stars (orange-shaded histogram).  }
\label{lievo}
\end{figure}

In \citet{fu2018} we select main sequence stars with UVES/VLT spectra from the $Gaia-ESO$ survey
and use their [$\alpha$/Fe] abundance to separate them into the Galactic thin and thick disc stars.
We find that the Galactic thin disc stars have higher Li abundance compared to those with similar \feh~ in the thick disc.
Fig. \ref{lievo} shows the results.
The fraction of stars with enriched-Li ($f_{A(Li)}>2.2$) is also higher in the thin disc than in the thick disc.
Our results suggest that the sources of the Galactic Li,
i.e. novae, AGB stars, Li-rich RGB stars, core-collapse supernovae, and cosmic ray, 
have different contributions in different Galactic components, 
which isipossibly linked to different star formation history.
Furthermore, in \citet{fu2018} we find an anti-correlation between A(Li) and [$\alpha$/Fe], 
which indicate that the $\alpha$ element producer core-collapse SNe are not the main source of Li,
and a correlation between A(Li) and s-process element abundance,
which can be explained by their common source AGB stars.

\bibliographystyle{aa}

\begin{thebibliography}{}

\bibitem[Bressan et al., (2012)]{bressan2012} Bressan, A., et al. \ 2012, \mnras, 427, 127

\bibitem[Taylor(2005)]{topcat} Taylor, M.~B.\ 2005, Astronomical Data Analysis Software and Systems XIV, 347, 29 

\bibitem[Boch \& Fernique (2014)]{aladin1} Boch, T., \& Fernique, P. \ 2014, 
         Astronomical Data Analysis Software and Systems XXIII. vol. 485, 2014, p.277

\bibitem[Bonnarel et al., (2000)]{aladin2} Bonnarel, F., et al. \ 2000, 
        Astronomy and Astrophysics Supplement Series, 143, 33

\bibitem[Fu et al., (2018)]{fu2018} Fu, X., et al. \ 2018, A\&A, 610A, 38F

\bibitem[Fu et al., (2015)]{fu2015} Fu, X., et al. \ 2015, \mnras, 452, 3256F

\bibitem[Lind et al., (2009)]{lind2009} Lind, K., et al. \ 2009, A\&A, 503, 545

\bibitem[Mucciarelli et al., (2011)]{mucciarelli2011} Mucciarelli, A., et al. \ 2011, \mnras, 412, 81

\bibitem[Aringer et al., (2016)]{Aringer2016} Aringer, B., et al. \ 2016, \mnras, 457, 3611A

\bibitem[Molaro et al., (2012)]{molaro2012} Molaro, P., et al. \ 2012, Mem. Soc. Astron. Ital. Suppl., 22, 233

\bibitem[Combes \& Wiklind (1998)]{cw1998} Combes, F., \& Wiklind, T., \ 1998, A\&A, 334, L81
        
\bibitem[Friedel et al., (2011)]{friedel2011} Friedel et al. \ 2011, \apj, 1, 37

\bibitem[Zhang et., (2018)]{csst} Zhang, K., et al. \ 2018, 2018 International Conference on Microwave and Millimeter Wave Technology 
\end{thebibliography}

\end{document}